\documentclass{PoS}
\usepackage{graphicx}
\usepackage{sidecap}

\title{Saving the Higgs Portal for Singlet Scalar Dark Matter}

\ShortTitle{Short Title for header}

\author{\speaker{Javier Quilis}
        IFT, UAM-CSIC\\
        E-mail: \email{javier.quilis@csic.es}}

\abstract{The Higgs-portal model with a singlet scalar Dark Matter particle is one of the simplest extensions to the Standard Model that can reproduce the relic density. Unfortunately this model is strongly constrained by direct and indirect DM detection, as well as by collider physics. Most of the parameter space is already ruled-out and the rest will be explored in the next future. We show that a simple extension of the DM sector with a second scalar singlet enables a substantial opening of the allowed window in the parameter space.}

\FullConference{EPS-HEP 2017, European Physical Society conference on High Energy Physics\\
		5-12 July 2017\\
		Venice, Italy}

\begin{document}

\section{Higgs Portal with a Real Singlet Scalar Dark Matter (SHP) }

The real singlet-scalar Higgs portal (SHP) model stands out as one of the most economical and popular scenarios. It simply consists of one extra real singlet scalar, $S$ (the DM particle), which is minimally coupled to the SM through interactions with the ordinary Higgs (the only ones allowed at the renormalizable level). The corresponding Lagrangian reads 
\begin{equation}
\mathcal{L}_{\rm SHP}=\mathcal{L}_{\rm SM}+\frac{1}{2}\partial_{\mu} S \partial^{\mu} S- \frac{1}{2}m_0^2 S^2-\frac{1}{2}\lambda_S |H|^2 S^2 -\frac{1}{4!}\lambda_{4} S^4 .
\label{HPlagr}
\end{equation}
Furthermore, a discrete symmetry $S\rightarrow -S$ has been imposed in order to ensure the stability of the DM particle. After electroweak (EW) symmetry breaking, $H^0=(v+h)/\sqrt{2}$, new terms appear, including a trilinear coupling between $S$ and the Higgs boson, $(\lambda_S v/2) h S^2$. 

\begin{SCfigure}
\centering 
\includegraphics[width=0.5\textwidth]{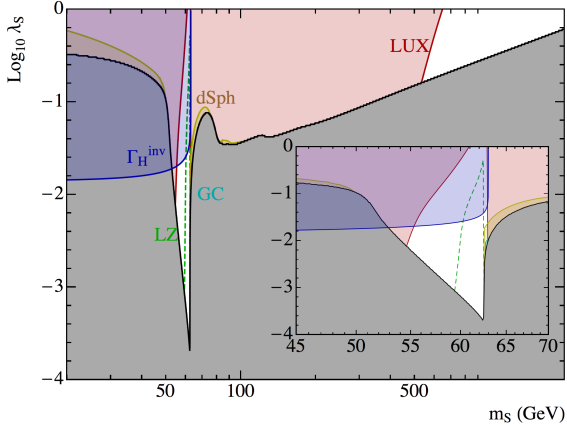}
\caption{
Excluded regions on the parameter space of the SHP model. The gray area is excluded since the relic density exceeds the Planck result. The blue area (labeled $\Gamma_H^{inv}$) is ruled out from the invisible Higgs width \cite{CMS:2016rfr}. The red area (LUX) is excluded by direct DM detection limits \cite{Akerib:2016vxi}. Yellow (dSph) and cyan (GC) areas are excluded by indirect detection constraints on the continuum spectrum of gamma-rays (from dwarf Spheroidal galaxies \cite{Ahnen:2016qkx}) and monochromatic gamma-ray lines (from the Galactic Centre \cite{Ackermann:2015lka}), respectively. The dashed green line represents the predicted reach of the future LZ detector. 
}
\label{fig:HPc}
\end{SCfigure}

Assuming that the $S-$particles are in thermal equilibrium in the early universe, the final DM relic density is determined by their primordial annihilation rate into SM-particles, that depends on just two parameters $\{m_S, \lambda_S\}$, where $m_S^2=m_0^2+\lambda_Sv^2/2$ is the physical $S-$mass after EW breaking. Fig.\,\ref{fig:HPc} 
shows the (black) line in the $\{m_S, \lambda_S\}$ plane along which
the relic abundance of $S$, $\Omega_Sh^2$, coincides with the Planck result $\Omega_{CDM}h^2=0.1198\pm 0.003$ at $2\sigma$ \cite{Ade:2015xua}. The (gray) region below is in principle excluded, as it corresponds to a higher relic density.

The model is subject to a number of experimental and observational constraints, which rule out large regions of the parameter space. These include limits from direct detection experiments, indirect searches, as well as collider bounds. We illustrate te effects of these limits in Fig.\,\ref{fig:HPc}. 

Next generation experiments such LZ \cite{Akerib:2015cja} will test completely the region of large DM masses and a large part of the narrow window at the Higgs-resonance.  The possibility of totally closing the Higgs-portal windows in the near future using complementary constraints from indirect detection has been analyzed, for example, in ref.~\cite{Feng:2014vea}.

In this proceeding we discuss the most economical modification of the conventional SHP model that could escape the present and future searches, proposed and analized in detail in our paper \cite{Casas:2017jjg} thas offering a viable (slightly different) Higgs-portal scenario if a positive detection does not occur. The model consist of the addition of a second singlet scalar in the dark sector, which opens up new annihilation and coannihilation channels.

\section{Higgs Portal with two Real Singlet Scalar Dark Matter (ESHP)}

The modification of the conventional SHP model that we consider consists simply of extending the DM sector with the addition of a second scalar.
 Denoting $S_1$, $S_2$ the two scalar particles, and imposing a global $Z_2$ symmetry ($S_1\rightarrow -S_1$, $S_2\rightarrow -S_2$) in order to guarantee the stability of the lightest one, the most general renormalizable Lagrangian reads
\begin{eqnarray}
\mathcal{L}_{\rm ESHP}&=&\mathcal{L}_{\rm SM}+\frac{1}{2}\sum_{i=1,2}\left[(\partial_{\mu} S_i)^2 -m_i^2 S_i^2-\frac{1}{12}\lambda_{i4}S_i^4\right]
-\frac{1}{6}\lambda_{13} S_1 S_2^3 -\frac{1}{6}\lambda_{31} S_1^3 S_2-\frac{1}{4}\lambda_{22} S_1^2 S_2^2
\nonumber\\
&&-\frac{1}{2}\lambda_1S_1^2|H|^2-\frac{1}{2}\lambda_2S_2^2|H|^2-\lambda_{12} S_1 S_2 \left(|H|^2 - \frac{v^2}{2}\right)\ ,
\label{HPextlagr}
\end{eqnarray}
where the subscript ESHP stands for ``extended singlet-scalar Higgs portal".
After EW breaking, there appear new terms, including trilinear terms between $S_{1,2}$ and the Higgs boson, such as $(\lambda_{12} v)\, h S_1S_2$. We have chosen $S_1, S_2$ to be 
 the final mass eigenstates (after EW breaking), with physical masses, $m_{S_i}^2 = m_i^2 + \lambda_i v^2/2$, thus the form of the last term in eq.(\ref{HPextlagr}). 
From now on, $S_1$ will represent the lightest mass eigenstate of the dark sector, and thus the DM particle. 

We will start by considering a scenario in which $\lambda_1$ is as small as possible ($\lambda_1 \sim \lambda_{12}^2/(4\pi)^2$), so, $\lambda_1$ can be neglected . On the other hand, we want to test the strength of the coupling $\lambda_{12}$, so, for that purpouse, we are setting the coupling $\lambda_2$ to this minimal magnitude $\lambda_2=\lambda_{12}^2/(4\pi)^2$ (we are fixing this value for $\lambda_2$ through the whole paper).  Therefore the only significant parameters to describe the DM physics are $m_{S_1}$, $m_{S_2}$, and $\lambda_{12}$.
For each  value of the DM mass, $m_{S_1}$, we are interested in finding out which combinations of $m_{S_2}$ and $\lambda_{12}$ lead to the correct relic density. 

\begin{figure}[t]
\centering 
\hspace*{-1ex}
\includegraphics[width=0.33\linewidth]{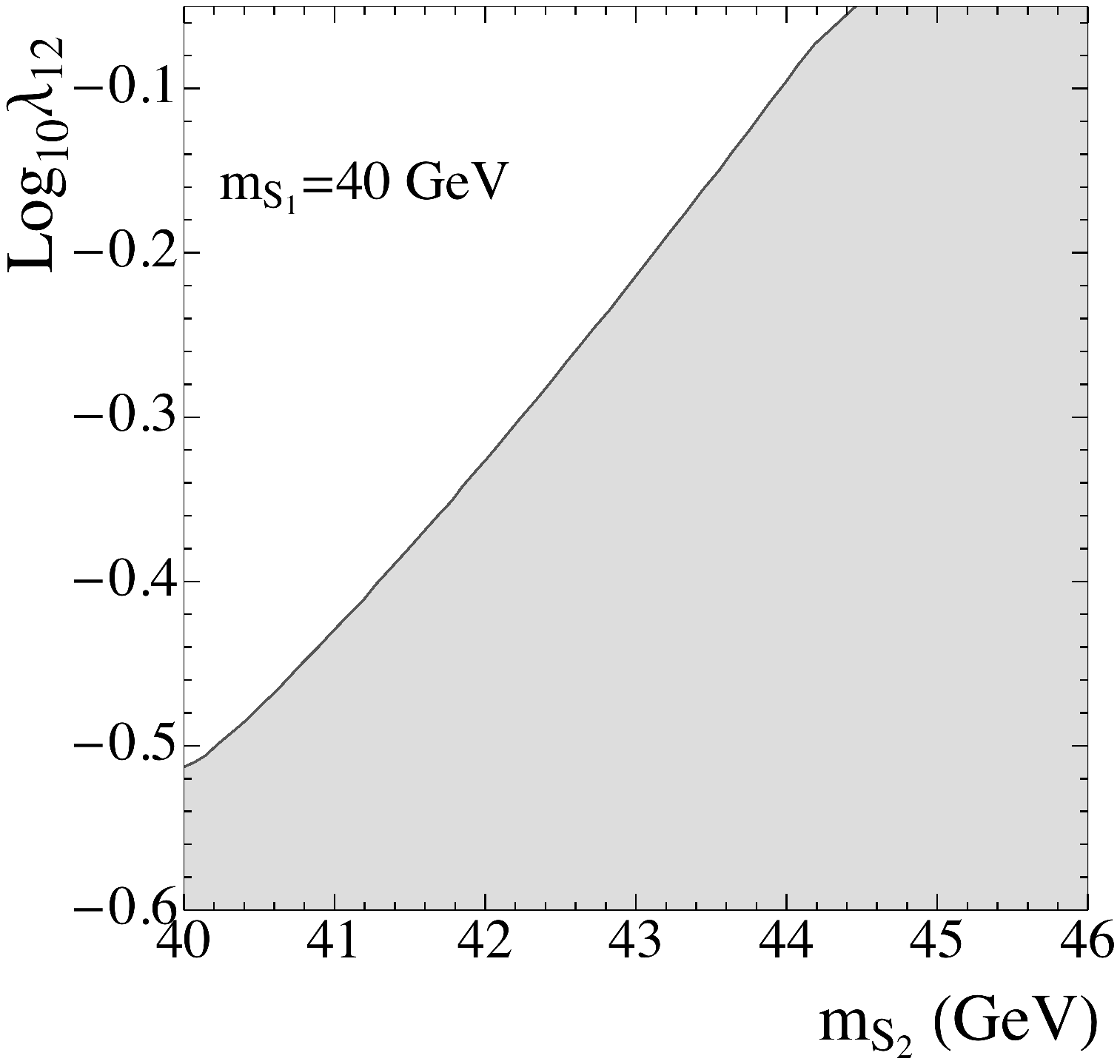}
\hspace*{-1ex}
\includegraphics[width=0.33\linewidth]{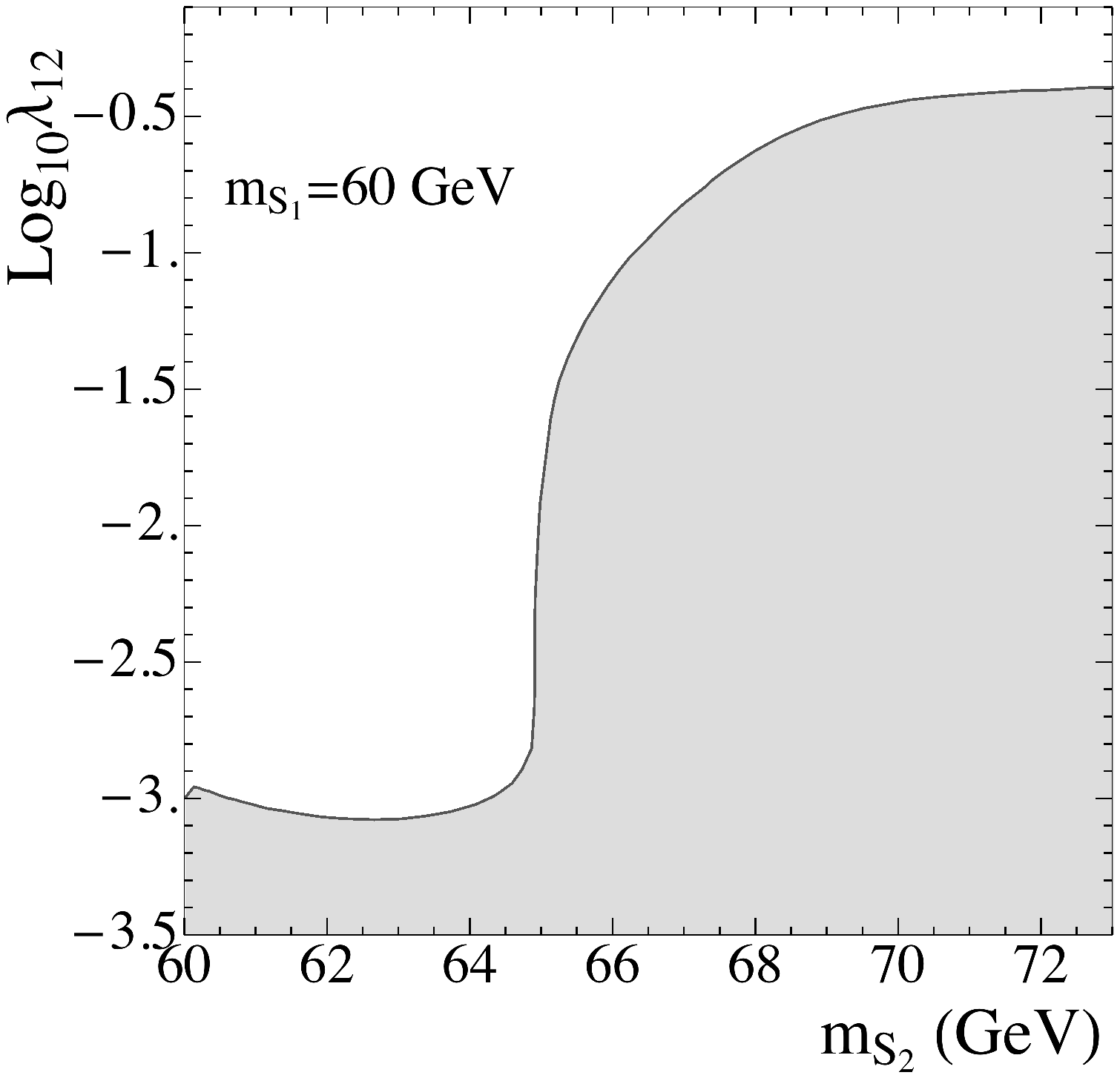}
\hspace*{-1ex}
\includegraphics[width=0.33\linewidth]{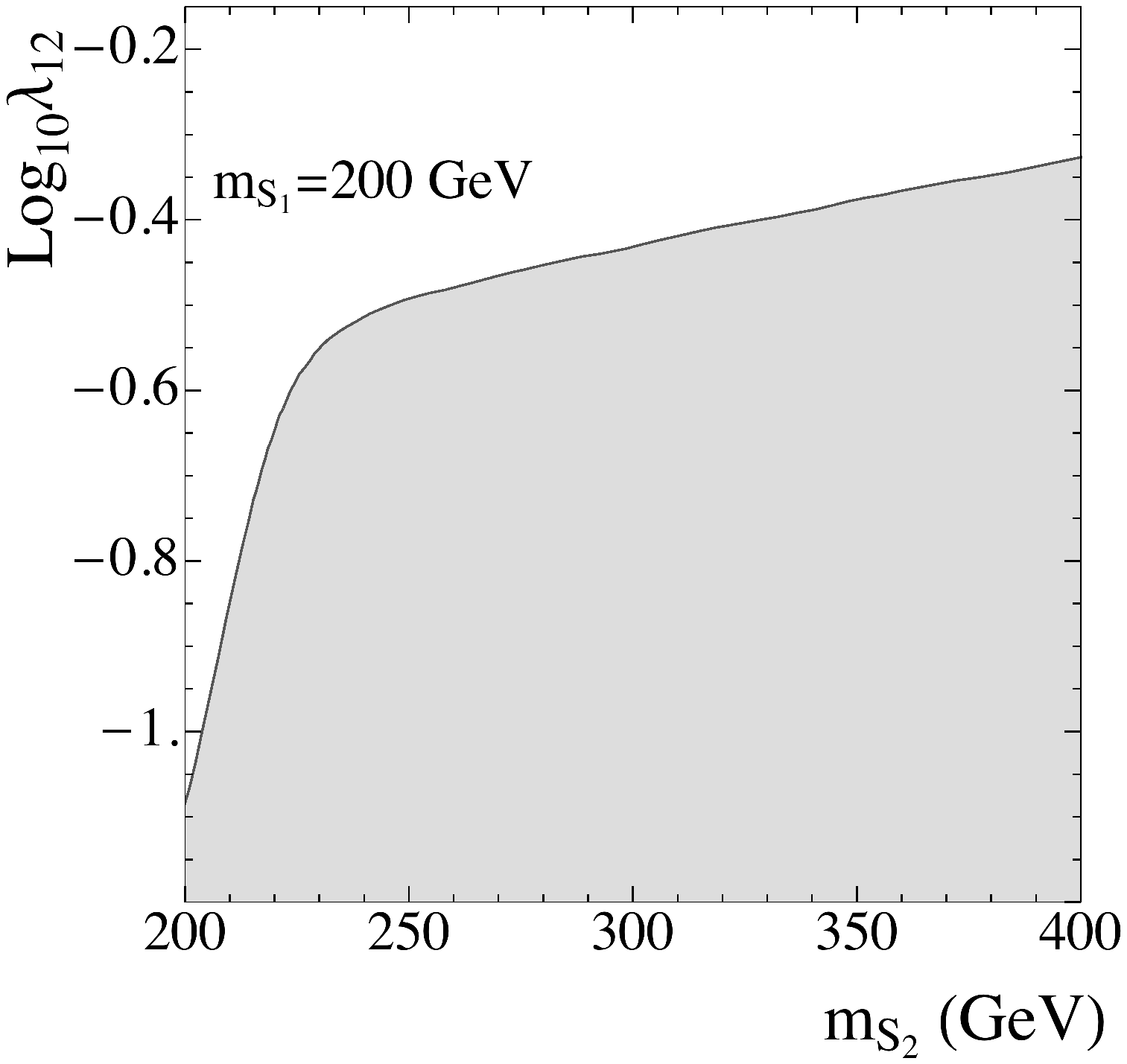}
\caption{
Range of values in the $\left\{\lambda_{12},\,m_{S_2}\right\}$ plane leading to the correct DM relic density
for three illustrative values of the DM mass: (from left to right) $m_{S_1}=40$ GeV, 60 GeV, and 200 GeV.
}
\label{fig:3examples}
\end{figure}

Fig.\,\ref{fig:3examples} shows the line along which the correct DM relic abundance is obtained 
for three representative cases, namely $m_{S_1}=40$, $60$, and $200$~GeV, i.e., below, around and above the Higgs resonance (left, middle and right panels, respectively).

Once we have checked that the copling $\lambda_{12}$ is enough to annihilate DM, we explore the parameter space of the ESHP model, applying all the contraints as we did for the SHP. Also the heavy scalar $S_2$ is unstable  and decays into $S_1$ (plus SM products). We will require that the decay occurs before Big Bang nucleosynthesis. Therefore, the lifetime of $S_2$ cannot be larger than 1s.

In order to facilitate the comparison of the model with the usual SHP, we have carried out a series of numerical scans, for fixed values of $\lambda_{12}$, in the three dimensional parameter space ($m_{S_1}, \,\lambda_1,\,m_{S_2}$), searching for points where $S_1$ is a viable candidate for dark matter\footnote{As already mentioned, we will set $\lambda_2$ at its lowest natural value, $\lambda_2=\lambda_{12}^2/(4 \pi)^2$. This is also the lower limits of $\lambda_1$ in the scans.} 

\begin{figure}[b]
\centering
\setlength{\tabcolsep}{0pt} 
\renewcommand{\arraystretch}{0} 
\begin{tabular}{c c c}
 $\lambda_{12}=$0.01  & $\lambda_{12}=$0.1  & $\lambda_{12}=1 $ \\ 
 \includegraphics[width=0.32\linewidth]{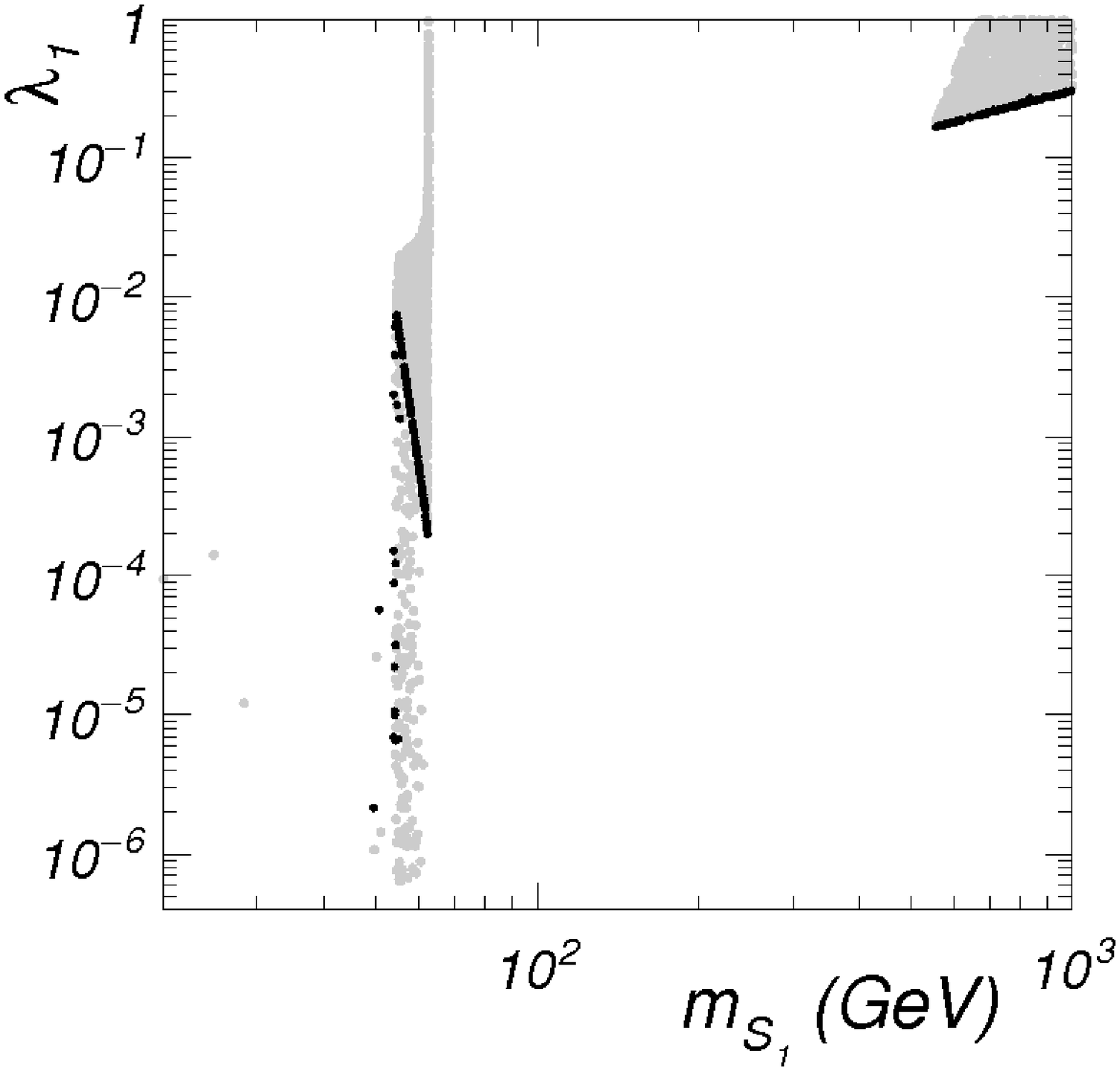} & \includegraphics[width=0.32\linewidth]{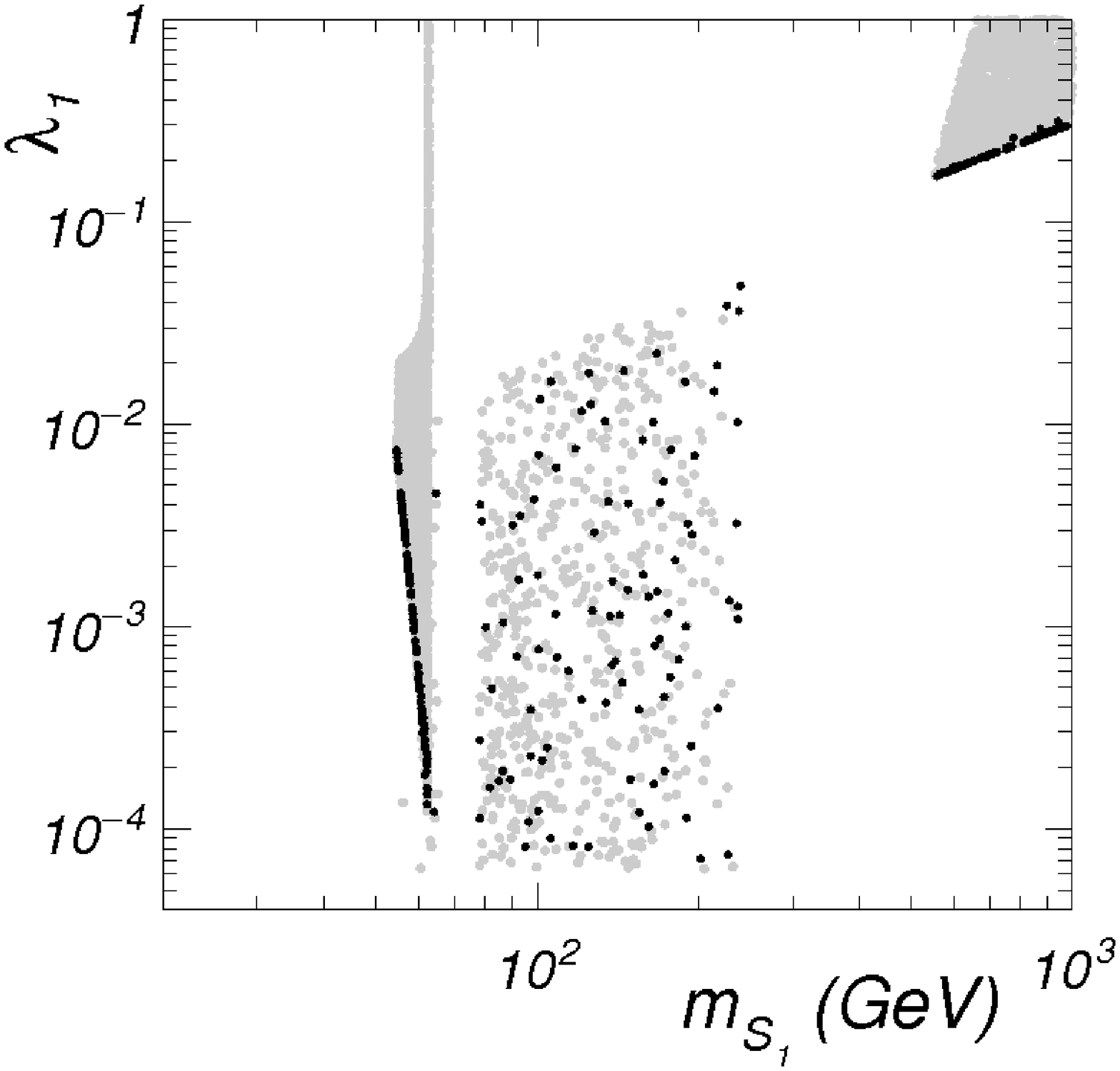} & \includegraphics[width=0.32\linewidth]{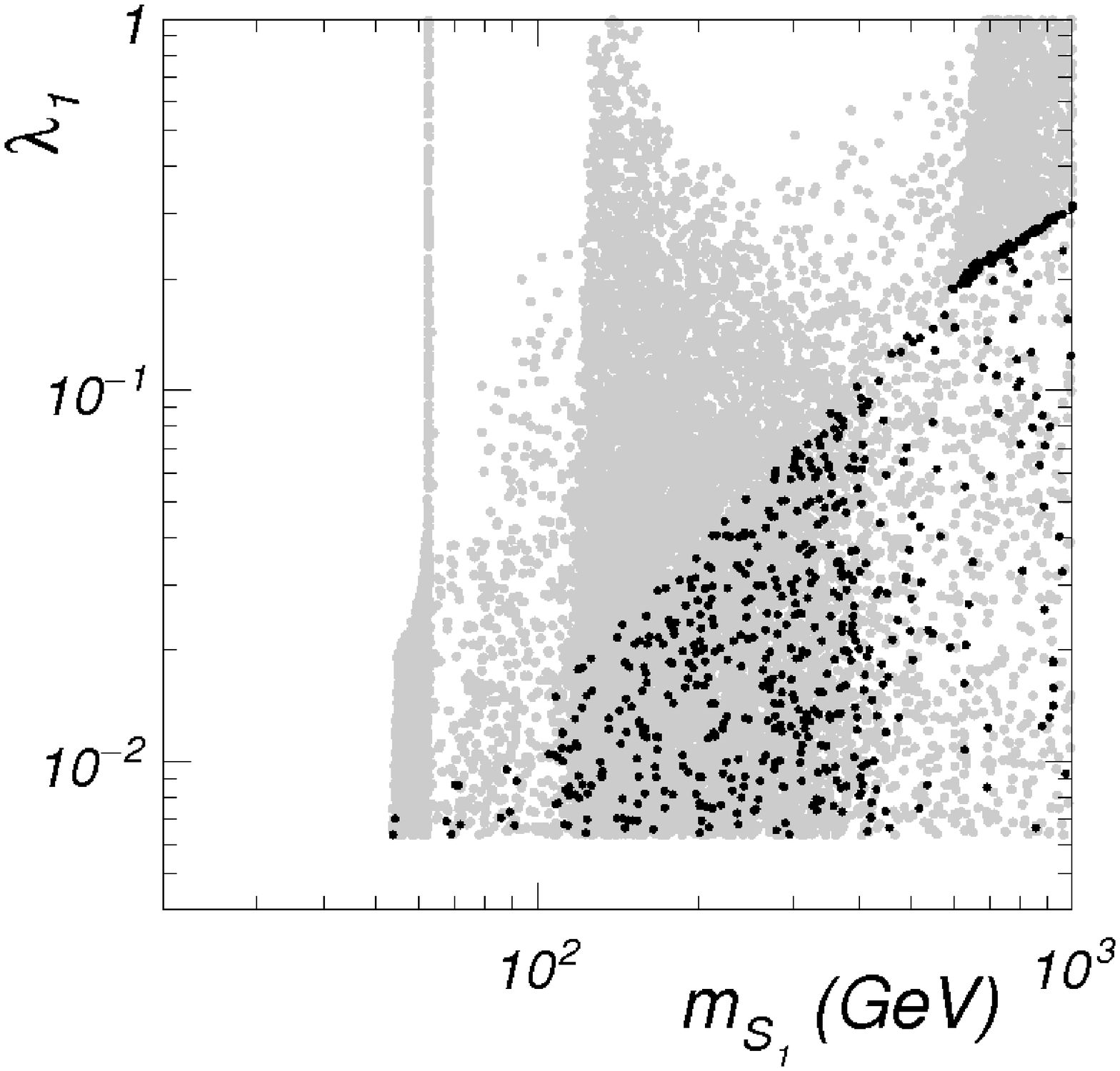} \\ 
\end{tabular}
\caption{
Effect of the experimental constraints in the $\{\lambda_1,\,S_{S_1}\}$ parameter space of the ESHP model. 
From left to right, we have fixed $\lambda_{12}=0.01,\,0.1,\,1$, and $\lambda_2=\lambda_{12}^2/(4\pi)^2$.
In all the plots, black (gray) points correspond to those where $\Omega h^2=0.119\pm 0.003$ ($\Omega h^2<0.116$). 
}
\label{fig:l1m1}
\end{figure}

We have represented the results of the scans in Fig.\,\ref{fig:l1m1} , where $\{m_{S_1},\,\lambda_1\}$ are plotted for fixed values of $\lambda_{12}$, gradually switching on the effect of the extra singlet in the model. The different experimental constraints are added in the plots.
In all the plots, black dots correspond to those in which the (thermal) relic abundance of $S_1$ matches the results from the Planck satellite, whereas grey points are those in which $S_1$ is a subdominant dark matter component. 

The results for the left figure ($\lambda_{12}=0.01$) resemble those of the usual SHP due to the smallness of $\lambda_{12}$. 
 Still, when these results are compared to Figure\,\ref{fig:HPc}, we observe a new (small) population of points at the Higgs resonance, with very small values of the coupling $\lambda_1$. This occurs when the masses of $S_2$ and $S_1$ are close enough so that coannihilation effects become important.  Away from the resonance region, the coannihilation effect is irrelevant due to the small size of $\lambda_{12}$ assumed here.

As we increase the value of $\lambda_{12}$, new areas of the parameter space become available. In the middle plot  of Fig.\,\ref{fig:l1m1}, ($\lambda_{12}=0.1$), we observe a region of black dots with masses $m_{S_1} \approx100-200$~GeV and a very small $\lambda_1$ coupling. These points have the correct relic abundance thanks to coannihilation effects, which requires $m_{S_1} \sim m_{S_2} $. 

When $\lambda_{12}=1$ (right pannel of Fig.\,\ref{fig:l1m1}), the effect of the DM annihilation in two Higgses, $S_1 S_1\rightarrow hh$, exchanging $S_2$ in $t-$channel, becomes more remarkable, as soon as it is kinematically allowed, i.e. for $m_{S_1} \geq m_h$. For smaller values of $m_{S_1} $ co-annihilation is still the main responsible for DM annihilation, thus requiring the $S_1, S_2$ masses to be closer. 

\begin{figure}[t!]
\centering
\setlength{\tabcolsep}{0pt} 
\renewcommand{\arraystretch}{0} 
\begin{tabular}{c c c}
 $\lambda_{12}=$0.01  & $\lambda_{12}=$0.1  & $\lambda_{12}=1 $ \\ 
 \includegraphics[width=0.32\linewidth]{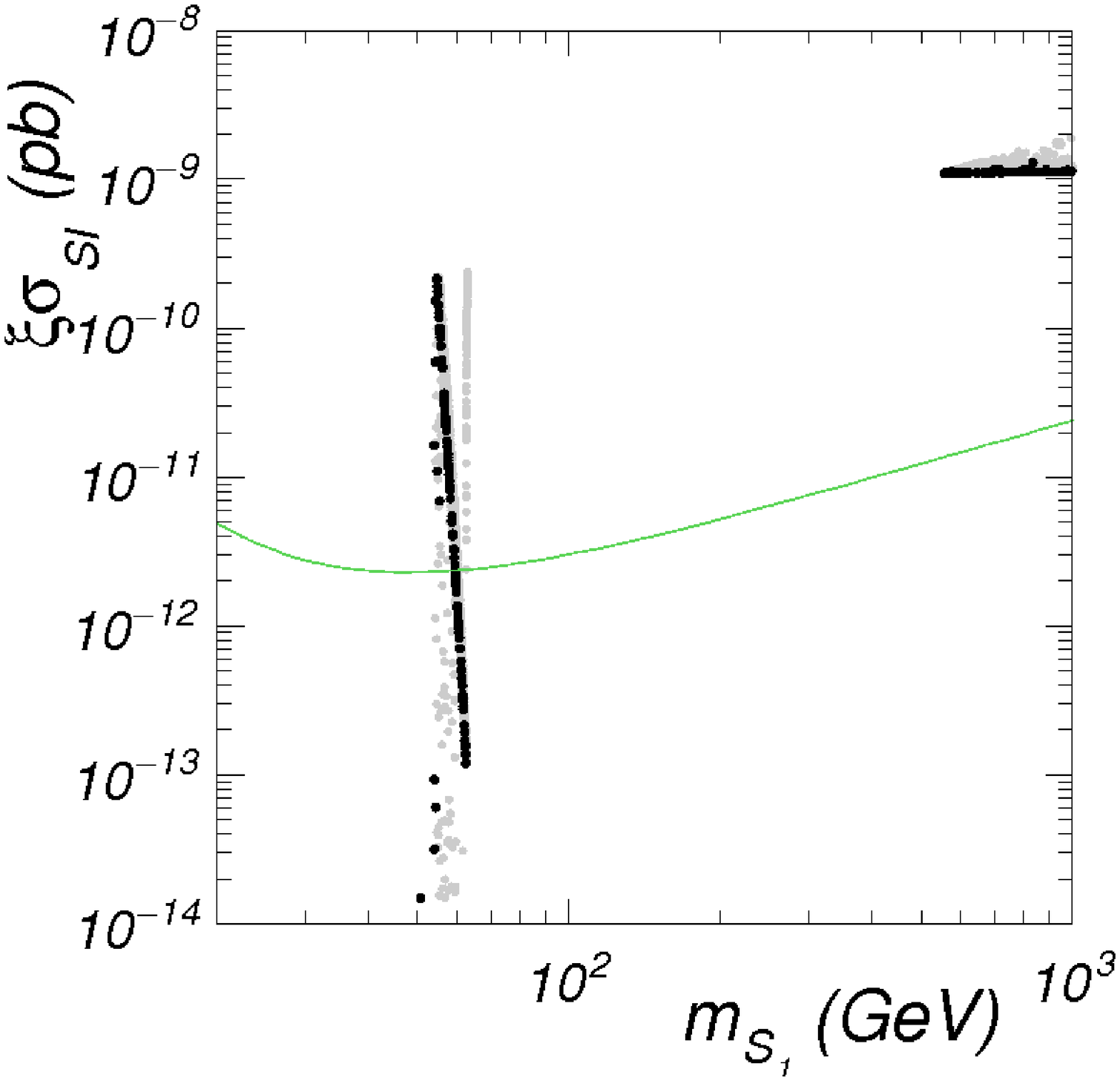} & \includegraphics[width=0.32\linewidth]{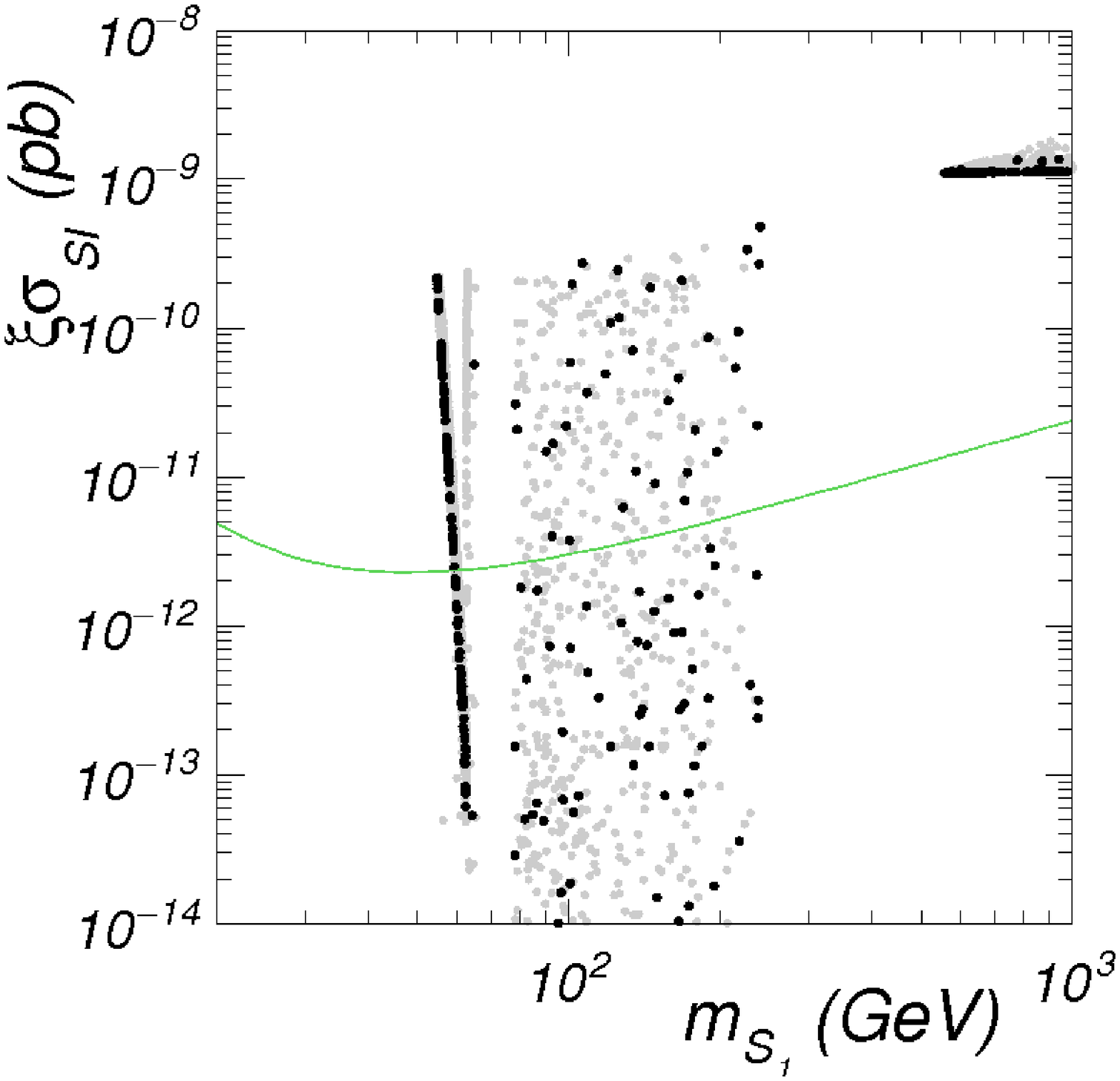} & \includegraphics[width=0.32\linewidth]{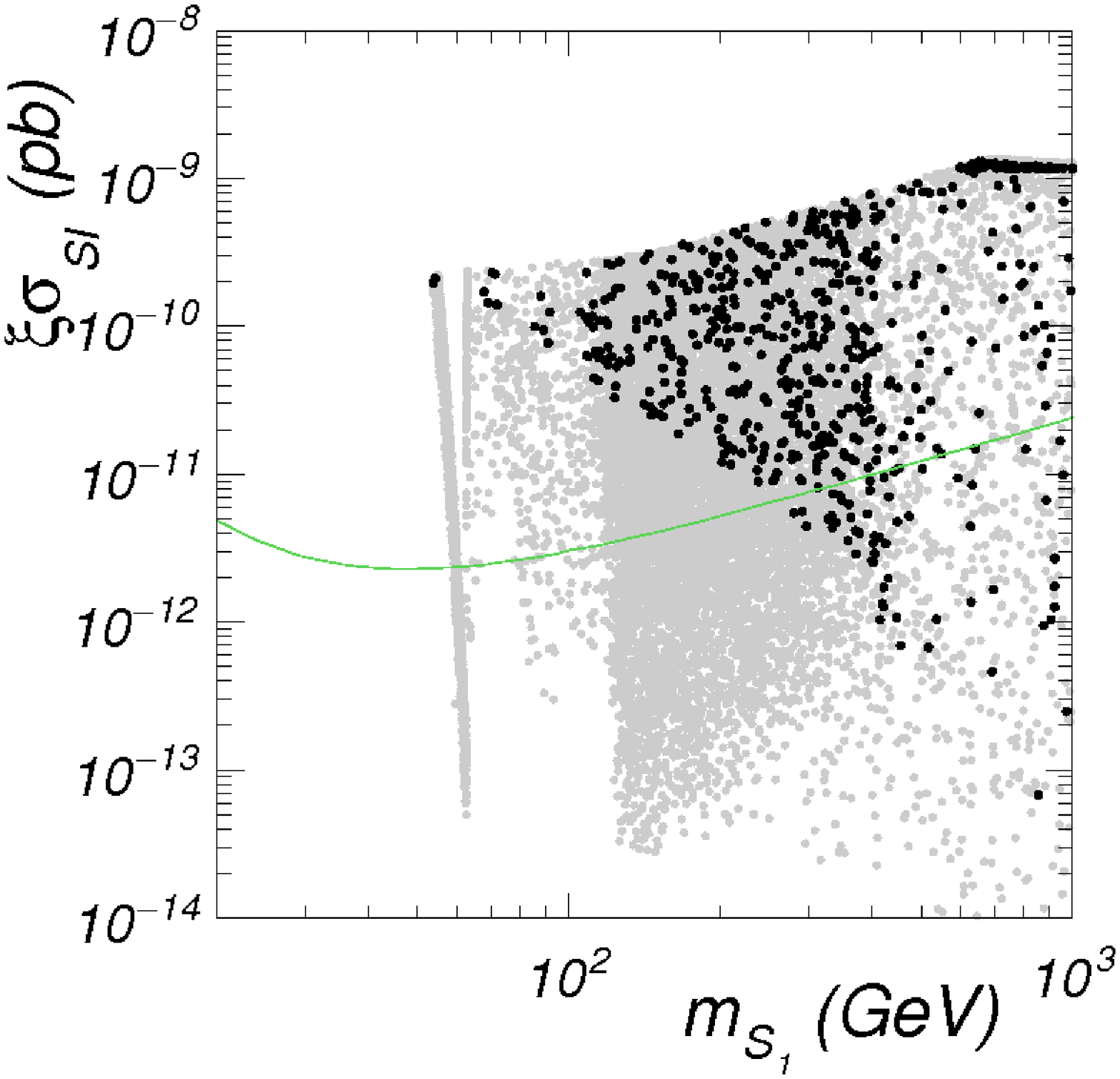} \\ 
\end{tabular}
\caption{Spin-independent scattering cross section of $S_1$ with protons as a function of its mass in the ESHP model. 
From left to right, we have fixed $\lambda_{12}=0.01,\,0,1$, and 1, respectively.}
\label{fig:lz}
\end{figure}

As in the case of the conventional SHP model, we expect future direct detection experiments (and in particular LZ) to be able to test large areas of the parameter space of our extended, ESHP, scenario. We represent in Fig.\,\ref{fig:lz} future LZ bounds, after all experimental constraints are applied. As we observe, although a large area of parameter space might be probed by these searches, there is a substantial region for which the prediction are beyond LZ sensitivity. 

\section{Conclusions}

Motivated by the appealing simplicity of this model, we have considered in this article a minimal extension (ESHP) that could evade detection. It consists  of the addition of an extra real singlet scalar field in the dark sector, coupled also in a minimal, renormalizable way. 

We show that the new annihilation and/or co-annihilation channels involving the extra singlet allow to reproduce the correct relic abundance, even if the usual interaction of the DM particle with the Higgs were arbitrarily small. This allows to easily avoid the bounds from direct and indirect DM searches.

\end{document}